\documentclass[preprint,preprint,prc,showpacs,preprintnumbers,
                unsortedaddress,amsmath,amssymb,floatfix]{revtex4}

\usepackage[T1]{fontenc}
\usepackage[utf8]{inputenc}
\usepackage{graphicx,color,rotating,pifont}
\usepackage{mathtools}
\usepackage{siunitx}
  
\usepackage{bm}
\usepackage{ae}

\usepackage{siunitx}
\usepackage{dcolumn}
\usepackage{txfonts}
\usepackage{tensor}
\usepackage{braket}
\usepackage{booktabs}
\usepackage{xspace}
\usepackage{mathrsfs}
\usepackage{amssymb} 
\usepackage{amsmath}
\usepackage{esvect}
\usepackage{diagbox} 

\usepackage{hyperref}

\DeclareSIUnit{\fm}{\femto\meter}

\newcommand{\nord}[1]{\ensuremath{\mathinner{\mathopen{:}{#1}\mathclose{:}}}}

\newcommand{\hw}{\ensuremath{\hbar\omega}}

\newcommand{\eMax}{\ensuremath{e_{\text{max}}}}

\newcommand{\beq}{\begin{equation}}
\newcommand{\eeq}{\end{equation}}
\newcommand{\beqn}{\begin{eqnarray}}
\newcommand{\eeqn}{\end{eqnarray}}
\newcommand{\bsub}{\begin{subequations}}
\newcommand{\esub}{\end{subequations}}
\newcommand{\bpm}{\begin{pmatrix}}
\newcommand{\epm}{\end{pmatrix}}

\begin{document}
 
\title{Quantum-number projected generator coordinate method for \nuclide[21]{Ne}   with a chiral two-nucleon-plus-three-nucleon  interaction}

 \author{W. Lin}
  \affiliation{School of Physics and Astronomy, Sun Yat-sen University, Zhuhai 519082, P.R. China}   

  \author{E. F. Zhou}
  \affiliation{School of Physics and Astronomy, Sun Yat-sen University, Zhuhai 519082, P.R. China}  

    \author{J. M. Yao}    \email{yaojm8@sysu.edu.cn}
  \affiliation{School of Physics and Astronomy, Sun Yat-sen University, Zhuhai 519082, P.R. China}

\author{H.~Hergert}    
\affiliation{Facility for Rare Isotope Beams, Michigan State University, East Lansing, Michigan 48824-1321, USA.}
\affiliation{Department of Physics \& Astronomy, Michigan State University, East Lansing, Michigan 48824-1321, USA.}


\begin{abstract}
In this paper, we report a study of the low-lying states of  deformed \nuclide[21]{Ne} within the framework of  quantum-number projected generator coordinate method (PGCM), starting from a chiral two-nucleon-plus-three-nucleon (NN+3N) interaction. The wave functions of states are constructed as a linear combination of a set of axially-deformed Hartree-Fock-Bogliubov (HFB) wave functions with different quadrupole deformations. These  HFB wave functions are projected onto different angular momenta and the correct neutron and proton numbers for \nuclide[21]{Ne}.  The results of calculations based on the effective Hamiltonians derived by normal-ordering the 3N interaction with respect to three different reference states, including the quantum-number projected HFB wave functions for \nuclide[20]{Ne}, \nuclide[22]{Ne}, and an ensemble of them with equal weights, are compared. This study serves as a key step towards ab initio calculations of odd-mass deformed nuclei with the in-medium GCM.
 
\end{abstract}

\maketitle


\section{Introduction}
Studying nuclear low-lying states, including energy spectra and electroweak transition strengths, is crucial for advancing our understanding of nuclear physics~\cite{Bohr:1998,Ring:1980}. It also plays a key role in exploring new physics at the high-precision frontier, such as nonzero electric dipole moments~\cite{Engel:2013PPNP,Arrowsmith-Kron:2023}, single-$\beta$ decay~\cite{Brodeur:2023}, and neutrinoless double-$\beta$ decay~\cite{Yao:2022PPNP}. Modeling the low-lying states of light to heavy atomic nuclei directly from the fundamental interactions between nucleons is of great interest for this purpose.  Compared to even-even nuclei, the low-lying states of odd-mass nuclei contain richer nuclear structure information because of the interplay of single-particle and collective motions, presenting a considerable challenge for nuclear theory. 
 
The generator coordinate method (GCM) provides an efficient and flexible framework to describe the wave function of a quantum many-body system, represented as a superposition of a set of nonorthogonal basis functions, such as Slater determinants, generated by continuously changing parameters called generator coordinates~\cite{Hill:1953,Griffin:1957}.   In nuclear physics, the quantum-number projected GCM (PGCM) has been extensively employed in studies of the energies and transition rates of low-lying states. See, for instance, Refs. \cite{Bender:2003SMF,Niksic:2011_PPNP,Sheikh:2021JPG}).  In the recent decade, the PGCM has been implemented into ab initio methods for atomic nuclei. This idea has given birth to a new generation of ab initio methods, including the no-core Monte Carlo shell model~\cite{Liu:2012PRC}, the in-medium 
generator coordinate method (IM-GCM)~\cite{Yao:2018PRC,Yao:2020PRL}  and perturbative PGCM with second-order perturbation theory~\cite{Frosini:2022_1,Frosini:2022_2,Frosini:2022_3}.

In this paper, we extend the PGCM for the low-lying states of an odd-mass deformed nucleus \nuclide[21]{Ne}, starting from a   Hamiltonian composed of  two-nucleon-plus-three-nucleon (NN+3N) interaction derived from chiral effective field theory (EFT). The PGCM has been extended for odd-mass nuclei based on different energy density functionals (EDFs)\cite{Kimura:2007,Kimura:2011,Bally:2014prl,Bally:2022_EPJA,Bally:2023_EPJA,Borrajo:2016vfz,Borrajo:2017,Borrajo:2018sqj,Zhou:2023_PRC}.  It is known that EDF-based PGCM approaches may suffer from spurious divergences and discontinuities~\cite{Bender:2009PRC,Duguet:2009PRC,Zhou:2023_PRC}. In this work, we examine that this Hamiltonian-based framework is free of those problems as the same interaction is applied to both the particle-hole and particle-particle channels when computing the energy overlaps of Hamiltonian kernels. Additionally, we compare the energy spectra of the low-lying states from the PGCM calculations using the effective Hamiltonian normal-ordered with respect to three different reference states, i.e., particle-number projected Hartree-Fock-Bogliubov (PNP-HFB) wave functions for \nuclide[20]{Ne}, \nuclide[22]{Ne}, and an ensemble of them with equal weights.

The article is arranged as follows. In  Sec.\ref{sec:framework}, we present the main formulas of PGCM for an odd-mass nucleus, including the generation of an effective Hamiltonian in the normal-ordering two-body (NO2B) approximation, and the construction of nuclear wave functions in the PGCM.   The results of calculations for \nuclide[21]{Ne} are presented in Sec.~\ref{sec:results}. A short summary and outlook are provided in Sec.~\ref{sec:summary}.


\section{The PGCM for an odd-mass nucleus}
\label{sec:framework}

\subsection{Nuclear Hamiltonian}
 We employ an intrinsic nuclear $A$-body Hamiltonian containing both $NN$ and $3N$ interactions, 
\begin{equation}
\label{Eq:H0}
\hat H_0 = \left(1-\frac{1}{A}\right) T^{[1]}+\frac{1}{A} T^{[2]}
+ \sum_{i<j} V_{ij}^{[2]} +  \sum_{i<j<k} W_{ijk}^{[3]},
\end{equation}
where the kinetic term is composed of one- and two-body pieces,
\beq 
T^{[1]} = \sum^A_{i=1} \frac{\mathbf{p}_i^2}{2 m_N}, \quad T^{[2]} =-\sum_{i<j} \frac{\mathbf{p}_i \cdot \mathbf{p}_j}{m_N},
\eeq 
with $m_N$ being the mass of nucleon and $\mathbf{p}_i$ the momentum of the $i$-th nucleon.

\begin{figure}[]
\centering
\includegraphics[width=\columnwidth]{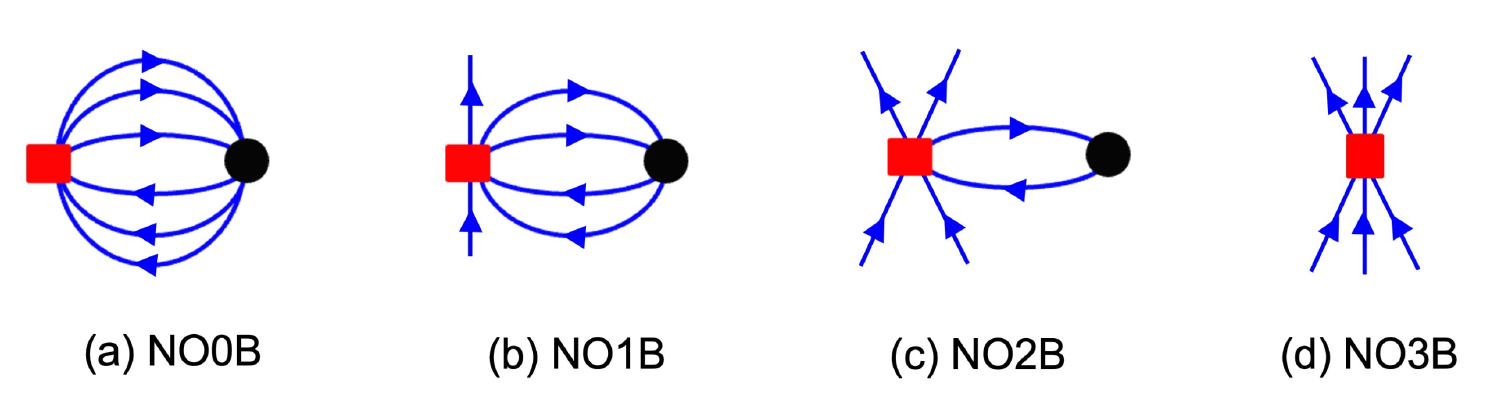}
\caption{Schematic illustration of the three-nucleon interaction $W$ (red squares), normal-ordered to (a) zero-body, (b) one-body, (c) two-body and (d) three-body terms with a reference state. The $n$-bdy density matrices $\gamma^{[n]}$ of the reference state, defined in (\ref{eq:density_correlated_state}), are represented with black circles. }
\label{fig:3N_NO}
\end{figure}

The above Hamiltonian is normal-ordered with respect to a symmetr-conserving reference state $\ket{\Psi}$,
and truncated up to NO2B terms. The resultant Hamiltonian $\hat {\cal H}_0$ in the NO2B approximation can be written as
\beqn 
\label{eq:H0_NO2B}
\hat {\cal H}_0
 = E_0 + \sum_{pq} f^{p}_{q} \nord{A^p_q}
+ \dfrac{1}{4}\sum_{pqrs}  \Gamma^{pq}_{rs}  \nord{A^{pq}_{rs}}.
\eeqn
 The strings of creation and annihilation operators are defined as 
\beq 
A^{pqr\ldots}_{stu\ldots} = a^\dagger_pa^\dagger_qa^\dagger_r\ldots a_u a_t a_s.
\eeq   
The expectation values of the normal-ordered operators, indicated by $\nord{A^{p\ldots}_{q\ldots}}$, with respect to the reference state are zero. The zero-body piece of the  ${\cal H}_0$ is just the energy of the reference state
\beqn
\label{H:0b}
 E_0\equiv \bra{ \Psi} \hat H_0\ket{ \Psi}
&=& \sum_{pq} \bar{t}^p_{q}\gamma^p_q 
+\dfrac{1}{4}\sum_{pqrs} \bar{v}^{pq}_{rs} \gamma^{pq}_{rs}  +\dfrac{1}{36}\sum_{pqrstu}w^{pqr}_{stu}\gamma^{pqr}_{stu} \,.
\eeqn
The matrix element of the normal-ordered one-body operator (NO1B) is given by
\beqn
\label{H:1b}
 f^{p}_{q}  &=& \bar{t}^{p}_{q}  +  \sum_{rs}  \bar{v}^{pr}_{qs}  \gamma^r_s
+\dfrac{1}{4}\sum_{rstu} w^{prs}_{qtu}\gamma^{rs}_{tu},
\eeqn  
and that of the NO2B operator,
\beqn
\label{H:2b}
 \Gamma^{pq}_{rs}  &=&\bar{v}^{pq}_{rs} + \sum_{tu} w^{pqt}_{rsu}\gamma^{t}_{u}\,.
\eeqn
The last terms in (\ref{H:0b}), (\ref{H:1b}) and (\ref{H:2b}) contributed by the 3N interaction are depicted schematically in Fig.~\ref{fig:3N_NO}(a), (b), and (c), respectively. Here,  we have introduced the density matrices of the (symmetry-conserving) correlated reference state $\ket{\Psi}$,
\bsub
\label{eq:density_correlated_state}
\beqn
\gamma^p_q &\equiv& \bra{ \Psi} A^{p}_{q}\ket{\Psi}\,,\\
\gamma^{pq}_{rs} &\equiv& \bra{ \Psi} A^{pq}_{rs}\ket{\Psi}\,,\\
\gamma^{pqr}_{stu} &\equiv& \bra{ \Psi} A^{pqr}_{stu}\ket{\Psi}\,.
\eeqn
\esub 
Static correlations within the reference state are encoded in the corresponding \emph{irreducible} density matrices  
\bsub
\label{eq:cumulants}
\beqn
\lambda^p_q &\equiv&  \gamma^p_q\,, \\
\lambda^{pq}_{rs} &\equiv& \gamma^{pq}_{rs}  - {\cal A}(\lambda^p_r\lambda^q_s)
 = \gamma^{pq}_{rs}  - \lambda^p_r\lambda^q_s +  \lambda^p_s\lambda^q_r\,,\\
\lambda^{pqr}_{stu} &\equiv& \gamma^{pqr}_{stu} - {\cal
A}(\lambda^p_s\lambda^{qr}_{tu}+\lambda^p_s\lambda^{q}_{t}\lambda^{r}_{u}) \,,
\eeqn
\esub
where the antisymmetrization operator ${\cal A}$ generates all possible
permutations (each only once) of upper indices and lower indices. For
a single-reference state,  the two-body and three-body irreducible densities $\lambda^{pq}_{rs}$ and $\lambda^{pqr}_{stu}$ vanish.

 The Hamiltonian $\hat {\cal H}_0$ is subsequently rewritten into the unnormal-ordered form as follows, 
\beqn
\label{eq:H0_UNO}
\hat {\cal H}_0 &=& \mathscr{E}_0 + \sum_{pq}\mathscr{F}^p_qA^p_q +\dfrac{1}{4} \sum_{pqrs}\mathscr{V}^{pq}_{rs}A^{pq}_{rs},
\eeqn
where the zero-body term is given by
 \bsub \beqn 
 \label{eq:UNH0_0b}
\mathscr{E}_0 &=&  E_0  -\sum_{pq}\left(f^p_q-\sum_{rs}\Gamma^{pr}_{qs}\gamma^r_s\right)\gamma^p_q
-\dfrac{1}{4}\sum_{pqrs}\Gamma^{pr}_{qs}\gamma^{pr}_{qs}\nonumber\\
&=&\dfrac{1}{36} \sum_{pqrstu} w^{pqr}_{stu}
 \left( \gamma^{pqr}_{stu} +36 \gamma^p_s\gamma^q_t\gamma^r_u - 18 \gamma^{pq}_{st}\gamma^r_u \right)\nonumber\\ 
 &=&\dfrac{1}{36} \sum_{pqrstu} w^{pqr}_{stu}
 \Bigg( 6\lambda^p_s\lambda^{q}_{t}\lambda^{r}_{u}-9\lambda^{pq}_{st}\lambda^r_u
  + \lambda^{pqr}_{stu} \Bigg).
\eeqn
The matrix elements of one-body read
\beqn 
 \label{eq:UNH0_1b}
\mathscr{F}^p_q &=&  f^p_q-\sum_{rs}\Gamma^{pr}_{qs}\gamma^r_s\nonumber\\
&=& \bar{t}^{p}_{q}   
+\dfrac{1}{4}\sum_{rstu} w^{prs}_{qtu}\Bigg(\lambda^{rs}_{tu}
- 2 \lambda^r_t\lambda^{s}_{u}\Bigg),
\eeqn 
and those of two-body terms
\beqn 
 \label{eq:UNH0_2b}
\mathscr{V}^{pq}_{rs} &=& \Gamma^{pq}_{rs}
=\bar{v}^{pq}_{rs} + \sum_{tu} w^{pqt}_{rstu}\lambda^{t}_{u}.
\eeqn
\esub

In this work, the reference state $\ket{\Psi}$ is chosen as a PNP-HFB state for  \nuclide[20]{Ne}, \nuclide[22]{Ne} and an ensemble of them with equal weights, which are labeled with {\tt magic-Ne20}, {\tt magic-Ne22}, and {\tt magic-ENO/EW}, respectively. The obtained effective Hamiltonians ${\cal H}_0$ are labeled as {\tt H0}. For comparison, we also derive the Hamiltonian without the $3N$ interaction term in (\ref{Eq:H0}), and this Hamiltonian is labeled as {\tt H0 (w/o 3N)}. The expressions for the one-, two-, and three-body density matrices of a spherical PNP-HFB state have been given in Ref.~\cite{Hergert:2017PS}. Subsequently, these Hamiltonians are employed into the PGCM calculations.

\subsection{Nuclear wave functions}

The wave functions of low-lying states for an odd-mass nucleus are constructed with the PGCM as follows, 
\begin{equation}
\label{eq:gcmwf}
\vert \Psi^{J\pi }_\alpha\rangle
=\sum_{c} f^{J\alpha\pi }_{c}  \ket{NZ J\pi; c},
\end{equation} 
Here, $\alpha$ distinguishes the states with the same angular momentum $J$, and the symbol $c$ is a collective label for the indices $(K,\kappa,\mathbf{q})$. The basis function with correct quantum numbers ($NZJ\pi$) is given by
\begin{equation}
\label{eq:basis}
\ket{NZ J\pi; c} 
=  \hat P^J_{MK} \hat P^N\hat P^Z \ket{\Phi^{\rm (OA)}_\kappa(\mathbf{q})},
\end{equation}
where $\hat P^{J}_{MK}$ and $\hat{P}^{N, Z}$  are projection operators that select components with the angular momentum $J$, neutron number $N$ and proton number $Z$~\cite{Ring:1980},
\begin{subequations}
\begin{align}
\label{eq:Euler_angles}
\hat P^{J}_{MK} &= \frac{2J+1}{8\pi^2}\int d\Omega D^{J\ast}_{MK}(\Omega) \hat R(\Omega),\\
\label{eq:PNP_gauge}
\hat P^{N_\tau} &= \frac{1}{2\pi}\int^{2\pi}_0 d\varphi_{\tau} e^{i\varphi_{\tau}(\hat N_\tau-N_\tau)}.
\end{align}
\end{subequations}
The operator $\hat P^J_{MK}$ extracts the component of angular momentum along the intrinsic axis $z$ defined by $K$. The Wigner $D$-function is defined as $D^{J}_{MK}(\Omega)\equiv\bra{JM}\hat R(\Omega)\ket{JK}=\bra{JM}e^{i\phi\hat J_z}e^{i\theta\hat J_y}e^{i\psi\hat J_z}\ket{JK}$, where $\Omega=(\phi, \theta, \psi)$ represents the three Euler angles. The $\hat N=\sum_k a^\dagger_k a_k$ is particle-number operator. The mean-field configurations $\ket{\Phi^{\rm (OA)}_\kappa(\mathbf{q})}$ for odd-mass nuclei can be constructed as one-quasiparticle excitations on even-even vacua~\cite{Ring:1980}, 
\begin{eqnarray}
\label{eq:odd-mass-wfs}
 \ket{\Phi^{\rm (OA)}_\kappa(\mathbf{q})}  =\alpha^\dagger_\kappa \ket{\Phi_{(\kappa)}(\mathbf{q})},\quad  \alpha_\kappa \ket{\Phi_{(\kappa)}(\mathbf{q})}=0,
\end{eqnarray} 
where $\ket{\Phi_{(\kappa)}(\mathbf{q})}$ is a HFB state with even-number parity labeled with the collective coordinate $\mathbf{q}$. The quasiparticle operators $(\alpha, \alpha^\dagger)$ are connected to single-particle operators $(a, a^\dagger)$ via the Bogoliubov transformation~\cite{Ring:1980}, 
\begin{eqnarray}
\label{eq:Bogoliubov_transformation}
\left(
\begin{array}{cc}
\alpha\\
\alpha^{\dag}\\
\end{array}
\right)=\left(
\begin{array}{cc}
U^\dag&V^{\dag}\\
V^T&U^T\\
\end{array}
\right)\left(
\begin{array}{cc}
a\\
a^\dag\\
\end{array}
\right),
\end{eqnarray}
where the $U, V$ matrices are determined by the minimization of  particle-number projected energy,
\beq 
\delta \frac{\bra{\Phi^{\rm (OA)}_\kappa(\mathbf{q})} \hat {\cal H} \hat P^N\hat P^Z \ket{\Phi^{\rm (OA)}_\kappa(\mathbf{q})}}
{\bra{\Phi^{\rm (OA)}_\kappa(\mathbf{q})}  \hat P^N\hat P^Z \ket{\Phi^{\rm (OA)}_\kappa(\mathbf{q})}}=0.
\eeq

Different from the recent study based on a covariant EDF in Ref.\cite{Zhou:2023_PRC}, where three different schemes were employed to construct the configurations for odd-mass nuclei within the BCS ansatz, in this work we obtain the configurations of one-quasiparticle states with odd-number parity self-consistently by simply exchanging the $k$-column of the $U$ and $V$ matrices in the HFB wave function~\cite{Ring:1980}:
\begin{equation}
(U_{pk},V_{pk}) \longleftrightarrow (V^\ast_{pk},U^\ast_{pk}),
\end{equation}
where the index $p=(\tau n\ell jm)_p\equiv (n_p, \xi_p)$ is a label for the spherical harmonic oscillator basis, and $k$ the label for a quasiparticle state. For simplicity, axial symmetry is assumed. In this case, quasiparticle states are labeled with quantum numbers $K^\pi$, where $K=|m_p|$ with $m_p$ being the projection of angular momentum $j_p$ along $z$-axis, and parity $\pi=(-1)^{\ell_p}$. The collective coordinate $\mathbf{q}$ is replaced with the dimensionless quadrupole deformation $\beta_2$,
\beq 
\beta_2=\frac{4\pi}{3AR^2}\bra{\Phi^{\rm (OA)}_\kappa(\mathbf{q})} r^2Y_{20}\ket{\Phi^{\rm (OA)}_\kappa(\mathbf{q})}.
\eeq 
The $U$ and $V$ matrices are determined from the HFB calculation within the scheme of variation after particle-number projection(VAPNP). For details, see, for instance, Ref.~\cite{Bally:2021_EPJA}. We note that the Kramer's degeneracy is lifted due to the breaking of time-reversal invariance in the self-consistent HFB calculation.
  
 The weight function $f^{J\alpha \pi}_{c}$ of the state (\ref{eq:gcmwf}) is determined by the variational principle, which leads to the following Hill-Wheeler-Griffin (HWG) equation~\cite{Hill:1953,Ring:1980},
\begin{eqnarray}
\label{eq:HWG}
\sum_{c'}
\Bigg[\mathscr{H}^{NZJ\pi}_{cc'}
-E_\alpha^{J }\mathscr{N}^{NZJ\pi }_{cc'} \Bigg]
f^{J\alpha\pi }_{c'}=0,
\end{eqnarray}
where the Hamiltonian kernel  and norm kernel are defined by
\begin{eqnarray}
\label{eq:kernel}
 \mathscr{O}^{NZJ\pi}_{cc'}
 &=&\bra{NZ J\pi; c}  \hat O \ket{NZ J\pi; c'} \nonumber\\ 
 &=&  \frac{2J+1}{8\pi^2} \int d\Omega D_{KK'}^{J\ast}(\Omega)
 \int_0^{2\pi}d\varphi_n \frac{e^{-iN\varphi_n}}{2\pi} 
  \int_0^{2\pi}d\varphi_p \frac{e^{-iZ\varphi_p}}{2\pi} \nonumber\\
&&\times \bra{\Phi^{\rm (OA)}_{\kappa}(\mathbf{q})}
    \hat O \hat R(\Omega)  e^{i\hat Z\varphi_p}e^{i\hat N\varphi_n} \ket{\Phi^{\rm (OA)}_{\kappa'}(\mathbf{q}')},
\end{eqnarray}
with the operator $\hat O$ representing $\hat {\cal H}$ and 1, respectively. The parity $\pi$ is defined by the quasiparticle configurations $\ket{\Phi^{\rm (OA)}_{\kappa}(\mathbf{q})}$.

The HWG equation (\ref{eq:HWG}) for a given set of quantum numbers $(NZJ)$ is solved  in the standard way  as discussed in Refs.\cite{Ring:1980,Yao:2010}. It is accomplished by diagonalizing the norm kernel $\mathscr{N}^{NZJ\pi }_{cc'}$ first. A new set of basis is constructed using the  eigenfunctions of the norm kernel with eigenvalue larger than a pre-chosen cutoff value to remove possible redundancy in the original basis. The Hamiltonian is diagonalized in this new basis. In this way, one is able to obtain the energies $E_\alpha^{J}$ and
the mixing weights $f^{J\alpha\pi}_{c}$ of nuclear states $\vert \Psi^{J\pi}_\alpha\rangle$. Since the basis functions $\ket{NZ J; c}$ are nonorthogonal to each other, one usually introduces the collective wave function $g^{J\pi}_\alpha(K, \mathbf{q})$ as below
\begin{equation}
\label{eq:coll_wf}
g^{J\pi}_\alpha(K, \mathbf{q})=\sum_{c'} (\mathscr{N}^{1/2})^{NZJ\pi}_{c,c'} f^{J\alpha\pi}_{c'},
 \end{equation}
 which fulfills the normalization condition. The distribution of $g^{J\pi}_\alpha(K, \mathbf{q})$ over $K$ and $\mathbf{q}$ reflects the contribution of each basis function to the nuclear state $\vert \Psi^{J\pi}_\alpha\rangle$.

\subsection{Evaluation of norm and Hamiltonian overlaps}
 The energy overlap is defined as the  ratio of Hamiltonian overlap to the norm overlap,
\beqn
E(\kappa\mathbf{q},\kappa'\mathbf{q}';g) 
&\equiv &\frac{\bra{\Phi^{\rm (OA)}_{\kappa}(\mathbf{q})}
    \hat {\cal H} \hat R(\Omega)  e^{i\hat Z\varphi_p}e^{i\hat N\varphi_n} \ket{\Phi^{\rm (OA)}_{\kappa'}(\mathbf{q}')} }{\bra{\Phi^{\rm (OA)}_{\kappa}(\mathbf{q})}
     \hat R(\Omega)  e^{i\hat Z\varphi_p}e^{i\hat N\varphi_n} \ket{\Phi^{\rm (OA)}_{\kappa'}(\mathbf{q}')}}\nonumber\\
 &=& \mathscr{E}
 +\sum_{pq}\mathscr{F}^p_q \tilde\rho^{p}_q(\kappa\mathbf{q},\kappa'\mathbf{q}';g) 
  +\dfrac{1}{4} \sum_{pqrs}\mathscr{V}^{pq}_{rs} 
  \tilde\rho^{pq}_{rs}(\kappa\mathbf{q},\kappa'\mathbf{q}';g),
\eeqn 
where $g$ stands for the set of parameters $\{\Omega,\varphi_n,\varphi_p\}$. The matrix elements of the mixed one-body densities and pairing tensors, hatted with the symbol $\sim$, are defined as
\beqn 
\tilde\rho^{p}_q(\kappa\mathbf{q},\kappa'\mathbf{q}';g) 
&\equiv&
\frac{\bra{\Phi^{\rm (OA)}_{\kappa}(\mathbf{q})}
    a^\dagger_p a_q \hat R(\Omega)  e^{i\hat Z\varphi_p}e^{i\hat N\varphi_n} \ket{\Phi^{\rm (OA)}_{\kappa'}(\mathbf{q}')} }{\bra{\Phi^{\rm (OA)}_{\kappa}(\mathbf{q})}
     \hat R(\Omega)  e^{i\hat Z\varphi_p}e^{i\hat N\varphi_n} \ket{\Phi^{\rm (OA)}_{\kappa'}(\mathbf{q}')}},\\
\tilde\kappa^{pq}(\kappa\mathbf{q},\kappa'\mathbf{q}';g)  
&\equiv&
\frac{\bra{\Phi^{\rm (OA)}_{\kappa}(\mathbf{q})}
    a^\dagger_p a^\dagger_q \hat R(\Omega)  e^{i\hat Z\varphi_p}e^{i\hat N\varphi_n} \ket{\Phi^{\rm (OA)}_{\kappa'}(\mathbf{q}')} }{\bra{\Phi^{\rm (OA)}_{\kappa}(\mathbf{q})}
     \hat R(\Omega)  e^{i\hat Z\varphi_p}e^{i\hat N\varphi_n} \ket{\Phi^{\rm (OA)}_{\kappa'}(\mathbf{q}')}},\\
\tilde\kappa_{rs}(\kappa\mathbf{q},\kappa'\mathbf{q}';g) 
&\equiv&
\frac{\bra{\Phi^{\rm (OA)}_{\kappa}(\mathbf{q})}
    a_s a_r \hat R(\Omega)  e^{i\hat Z\varphi_p}e^{i\hat N\varphi_n} \ket{\Phi^{\rm (OA)}_{\kappa'}(\mathbf{q}')} }{\bra{\Phi^{\rm (OA)}_{\kappa}(\mathbf{q})}
     \hat R(\Omega)  e^{i\hat Z\varphi_p}e^{i\hat N\varphi_n} \ket{\Phi^{\rm (OA)}_{\kappa'}(\mathbf{q}')}}.
\eeqn 
The matrix elements of the mixed two-body density are determined by the generalized Wick theorem~\cite{Balian:1969},
\beqn 
 \tilde\rho^{pq}_{rs}(\kappa\mathbf{q},\kappa'\mathbf{q}';g)
&\equiv&
\frac{\bra{\Phi^{\rm (OA)}_{\kappa}(\mathbf{q})}
    a^\dagger_p a^\dagger_q a_sa_r \hat R(\Omega)  e^{i\hat Z\varphi_p}e^{i\hat N\varphi_n} \ket{\Phi^{\rm (OA)}_{\kappa'}(\mathbf{q}')} }{\bra{\Phi^{\rm (OA)}_{\kappa}(\mathbf{q})}
     \hat R(\Omega)  e^{i\hat Z\varphi_p}e^{i\hat N\varphi_n} \ket{\Phi^{\rm (OA)}_{\kappa'}(\mathbf{q}')}}\nonumber\\
     &=& \tilde\rho^{p}_r  \tilde\rho^{q}_s 
     -\tilde\rho^{p}_s  \tilde\rho^{q}_r
     + \tilde\kappa^{pq} \tilde\kappa_{rs}. 
\eeqn 

With the above relation, the energy overlap can be rewritten as below,
\beqn
E(\kappa\mathbf{q},\kappa'\mathbf{q}';g) 
 &=& \mathscr{E}
 +\sum_{pq}\mathscr{F}^p_q \tilde\rho^{p}_q 
+ \dfrac{1}{2} \sum_{pq} \Bigg(\tilde\Gamma^p_q\tilde\rho^p_q + \tilde\Delta^{pq}\tilde\kappa_{pq} \Bigg),
\eeqn 
where the matrix elements of the mixed particle-hole field $\tilde\Gamma$ and particle-particle field $\tilde\Delta$ are defined as
 \beq
 \label{eq:mix_fields}
\tilde\Gamma^p_q \equiv \sum_{pqrs} \mathscr{V}^{pr}_{qs}  \tilde\rho^r_s, \quad
\tilde\Delta^{pq} \equiv  \frac{1}{2} \sum_{rs} \mathscr{V}_{pq}^{rs} \tilde\kappa^{rs}.
\eeq
It is efficient to compute the energy overlap directly in the $J$-coupled scheme. 

\begin{itemize}
    \item The contribution of the one-body term is simply given by 
 \beqn
    E^{(1B)} =   \sum_{pq} \delta_{j_p j_q} \hat j_p  \mathscr{F}^0_{(qp)}\tilde\rho_{(qp)00}, 
    \eeqn 
    where $\hat j_p\equiv \sqrt{2j_p+1}$. The reduced matrix element is defined as $\mathscr{F}^0_{qp}=\bra{q}|\mathscr{F}_{0}|\ket{p}$, and the one-body density operator with the two angular momenta coupled to zero~\cite{Yao:2018PRC}  
    \beq 
    \hat\rho_{(qp)00}\equiv \frac{ \left[a_{q}^{\dagger} \tilde{a}_{p}\right]_{00} }{\sqrt{2 j_q+1}} \delta_{\xi_q \xi_p}
   \eeq   
with $\tilde a_{nljm}\equiv (-1)^{j+m}a_{nlj-m}$.  
  \item The energy by the two-body term consists of $pp$ term
   \beqn
   \label{eq:pairing_energy_J}
   E^{(2B)}_{pp}  =  -\dfrac{1}{4}\sum_{abcd, L}  \mathscr{V}^L_{(ab) (cd)}  \sum_{M_L} (-1)^{L-M_L}
       \tilde\kappa^{(01)}_{(ab)LM_L}    \tilde\kappa_{(cd)L-M_L}^{(10)} 
    \eeqn    
   and $ph$ term 
 \beqn
   \label{eq:ph_energy_J}
 E^{(2B)}_{ph} &=&     \dfrac{1}{2} \sum_{abcd,L}   
      \mathscr{V}^L_{(a\bar b) (c\bar d)} 
      \sum_{M_L}   \tilde\rho_{(ba)LM_L}  \tilde\rho^\dagger_{(dc)L-M_L} 
    \eeqn 
    where the $J$-coupled mixed density and pairing density are defined as,
\bsub \beqn
  \tilde\rho_{(ba)LM_L}
  &=&  
    \sum_{m_am_b }  s_b
    \langle j_am_a j_b -m_b\vert LM_L\rangle 
    \tilde\rho^a_b,\\
    \tilde\rho^\dagger_{(dc)L-M_L} 
  &=&  
    \sum_{m_cm_d }  s_d
    \langle j_cm_c j_d -m_d\vert L-M_L\rangle 
    (\tilde\rho^c_d)^\dagger,\\
    \tilde\kappa^{(01)}_{(ab)LM_L}  
  &=&\sum_{m_am_b }  
    \langle j_am_a j_b m_b\vert LM_L\rangle 
    \tilde\kappa^{ab},\\
  \tilde\kappa^{(10)}_{(cd)L-M_L}  
  &=&(-1)^{L+M_L}\sum_{m_cm_d }  
    \langle j_cm_c j_d m_d\vert LM_L\rangle 
    (\tilde\kappa_{cd})^\dagger.
    \eeqn
\esub 
Here, we introduce the symbol $s_b\equiv (-1)^{j_b-m_b}$. The symmetry of Clebsch–Gordan coefficient $ \langle j_am_a j_b -m_b\vert LM_L\rangle$ implies the relation $ \tilde\rho_{(ba)LM_L}
=(-1)^{L-(j_a+j_b)+1} \tilde\rho_{(ab)LM_L}$. The $ph$-type two-body interaction matrix elements in the $J$-coupled form are related to those of $pp$-type by Pandya transformation~\cite{Pandya:PR1956},
	\beqn
	\mathscr{V}^J_{(i\bar j) (k\bar l)}
	=-\sum_L \hat L^2    \left\{\begin{array}{ccc}
   j_i & j_j & J \\
   j_k & j_l  & L 
    \end{array}\right\}  \mathscr{V}^L_{(il) (kj)},
	\eeqn
where the unnormalized $pp$-type two-body matrix elements in the $J$-coupled form are related to those in $M$-scheme as follows
\beqn
\mathscr{V}^{J}_{(ij)(kl)}
&=&\sum_{m_i m_jm_km_l} \braket{ j_im_i j_jm_j| JM} 
\braket{ j_km_k j_lm_l | JM}  \mathscr{V}^{ij}_{kl}.
\eeqn

\end{itemize}
The norm overlap of the HFB wave functions with odd-number parity is computed with the Pfaffian formula in Ref.~\cite{Avez:2012PRC}.  
  
\section{Results and discussion}
\label{sec:results}

\subsection{Effective Hamiltonians}

In this work, the $NN$ interaction $V_{ij}^{(2)}$ in Eq.(\ref{Eq:H0}) is chosen as the chiral N\textsuperscript3LO interaction by \citet{Entem:2003PRC}, denoted as "EM". We utilize the free-space SRG~\cite{Bogner:2010PPNP} to evolve the EM interaction to a resolution scale of $\lambda=1.8$ fm$^{-1}$. The $3N$ interaction $W_{ijk}^{(3)}$ is directly constructed with a cutoff of $\Lambda=2.0$ fm$^{-1}$. The Hamiltonian is referred to as EM$\lambda$/$\Lambda$, i.e., EM1.8/2.0, which was fitted to $NN$ scattering phase shifts, the binding energy of \nuclide[3]{H} ,
and the charge radius of \nuclide[4]{He}. See Ref.\cite{Hebeler:2011PRC} for details. For the 3N interaction, we discard all matrix elements involving states with $e_1+e_2+e_3>14$, where $e_i=2n_i+\ell_i$ denotes the number of harmonic oscillator major shells for the $i$-th state. The maximal value of $e_i$ is labeled with $\eMax$, and the frequency of the harmonic oscillator basis is chosen as $\hbar\omega=20$ MeV. In this work,  $\eMax=6$, and $\hw=20$ MeV are employed. Starting from the chiral NN+3N interaction, we produce three sets of effective Hamiltonians  labeled as {\tt magic-Ne20}, {\tt magic-Ne22}, and {\tt magic-ENO/EW}, respectively. These Hamiltonians are generated by normal-ordering the $3N$ interaction with respect to the reference states of spherical PNP-HFB states for \nuclide[20]{Ne}, \nuclide[22]{Ne}, and their ensemble with equal weights, respectively. The residual normal-ordered three-body term, c.f. Fig.~\ref{fig:3N_NO}(d), is neglected.  Table~\ref{table:Hamiltonian} lists the expectation value of each term in the three types of effective Hamiltonians ${\cal H}_0$ in (\ref{eq:H0_UNO}) with respect to the corresponding reference state.  One can see  that in the case without the $3N$ interaction, the unnormal-ordering form of the Hamiltonian ${\cal H}_0$  returns back to the original Hamiltonian $\hat H_0$.

The relative contribution of each term in different effective Hamiltonians to the energy is compared in Tab~\ref{table:Hamiltonian}. The contribution of the $3N$ interaction to energy, c.f. Fig.~\ref{fig:3N_NO}(a), is given by 
\beqn 
E^{(3)}_0
&=&\frac{1}{36}\sum_{pqrstu}w^{pqr}_{stu}\Bigg(
6\lambda^p_s\lambda^{q}_{t}\lambda^{r}_{u}+9\lambda^p_s\lambda^{qr}_{tu}+\lambda^{pqr}_{stu}\Bigg). 
\eeqn 
Comparing the $E_0$ value in the third row of Tab~\ref{table:Hamiltonian}, labeled by {\tt Ne20} with the $E_0$  value in the last row, labeled by {\tt Ne20 (w/o 3N)},  one finds the contribution of the $3N$ interaction to the energy $E^{(3)}_0=80.338$ MeV.    On the other hand, the zero-point energy $\mathscr{E}_0$ in (\ref{eq:UNH0_0b}) of the unnormal-ordered Hamiltonian in the first row 
 \beqn  
\mathscr{E}_0  
&=&\dfrac{1}{36} \sum_{pqrstu} w^{pqr}_{stu}
 \Bigg( 6\lambda^p_s\lambda^{q}_{t}\lambda^{r}_{u}-9\lambda^{pq}_{st}\lambda^r_u
  + \lambda^{pqr}_{stu} \Bigg) 
\eeqn
is 50.093 MeV. Their difference gives 
\beq 
\frac{1}{2}\sum_{pqrstu} w^{pqr}_{stu}\lambda^{pq}_{st}\lambda^r_u
=30.245\quad {\rm MeV}.
\eeq 
Since the term depending on $\lambda^{pqr}_{stu}$ is much smaller than the other terms, we drop this term out and find the term, 
\beq 
\frac{1}{6}\sum_{pqrstu}w^{pqr}_{stu} 
\lambda^p_s\lambda^{q}_{t}\lambda^{r}_{u}
= 65.215\quad {\rm MeV},
\eeq 
which depends solely on the one-body density, provides the predominant contribution to energies $E^{(3)}_0$ and $\mathscr{E}_0$. It implies that the terms depend on higher-order of irreducible densities $\lambda$ are less important. Subsequently,  we carry out PGCM calculations for low-lying states of \nuclide[21]{Ne} using the above effective Hamiltonians.

\begin{table}[] 
 \tabcolsep=16pt
\caption{The expectation value (in MeV)  of each term in the three different effective Hamiltonians ${\cal H}_0$ with resect to corresponding reference state.  }  
\begin{tabular}{ccccc}
  \toprule
      Interactions    & $E_0$    &  $\langle {\cal F} \rangle$ &  $\langle {\cal V} \rangle$   & $\mathscr{E}_0$ \\
      \hline  
 {\tt magic-Ne20}      &  $-96.931$      & 211.205   &  $-358.229$ &  50.093  \\
 {\tt magic-ENO/EW}   & $-101.781$    &    225.067       &  $-381.555$  &  54.706  \\ 
 {\tt magic-Ne22}    &  $-109.034$   &  242.241      & $-408.614$    &  57.339 \\
 \midrule
 {\tt magic-Ne20(w/o 3N)}   &  $-177.269$      & 506.122   &  $-683.391$ &  0  \\
   \bottomrule
\end{tabular}
\label{table:Hamiltonian}
\end{table}

\begin{figure}[]
\centering
\includegraphics[width=7.2 cm]{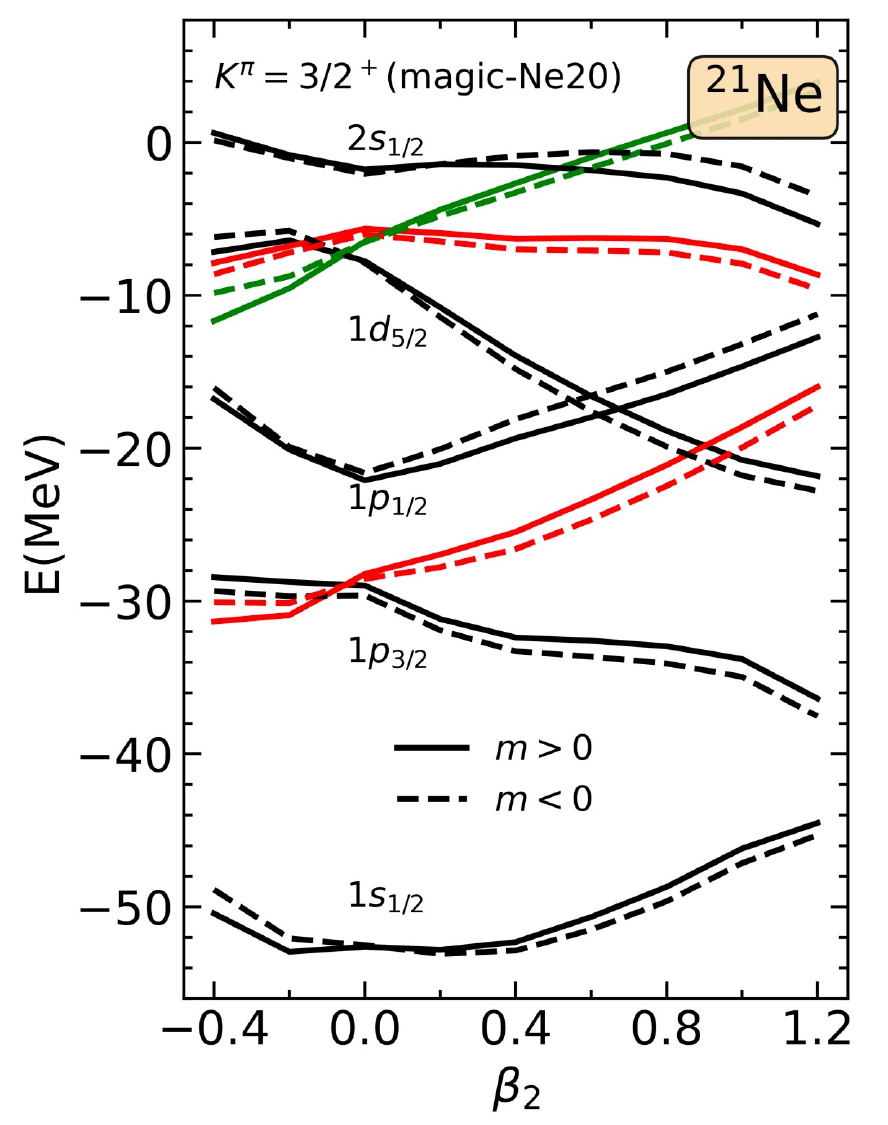}
\caption{ The effective single-particle energies of neutron states  with $m>0$ (solid lines) and $m<0$ (dashed lines) as a function of quadrupole deformation $\beta_2$ from the PNP-HFB (VAPNP) calculation for the HFB states with $K^\pi=3/2^+$ using the effective Hamiltonians {\tt magic-Ne20}.
}
\label{fig:SPE_interaction_A20}
\end{figure}  
 
\begin{figure}[]
\centering
\includegraphics[width=\columnwidth]{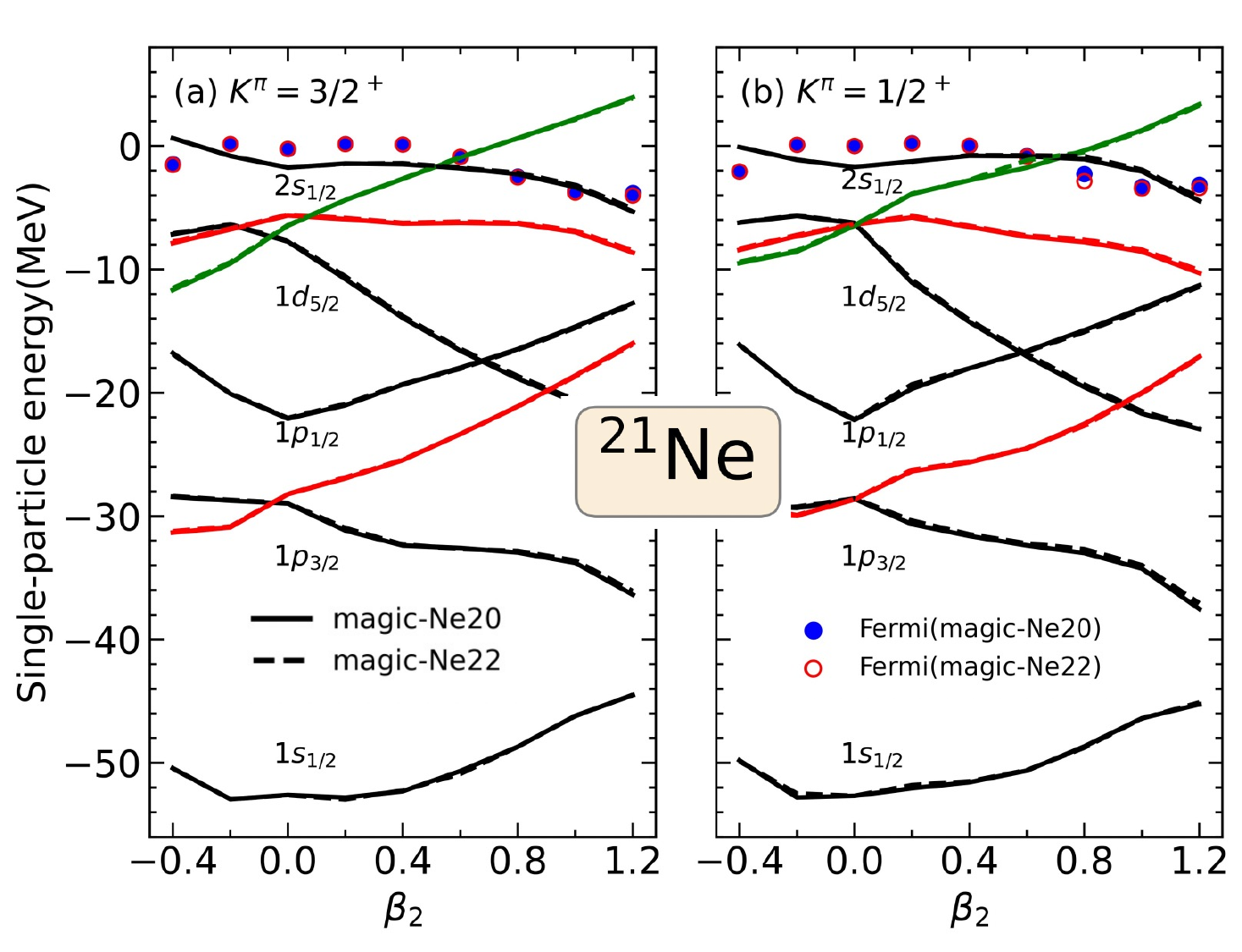}
\caption{ The effective single-particle energies of neutron states (with $m>0$) from the PNP-HFB (VAPNP) calculation for the HFB states with $K^\pi=3/2^+$ (a) and $K^\pi=1/2^+$ (b), where the effective Hamiltonians {\tt magic-Ne20} (solid) and {\tt magic-Ne22} (dashed lines) are employed, respectively. The Fermi energies are indicated with dots.
}
\label{fig:SPE_interaction_A20_A22}
\end{figure}  

Both Fig.~\ref{fig:SPE_interaction_A20} and Fig.~\ref{fig:SPE_interaction_A20_A22} show the change of the effective single-particle energies (ESPEs) with the quadrupole deformation $\beta_2$  from the PNP-HFB (VAPNP) calculation for the HFB states with different $K^\pi$, where the PNP is carried before variation. The ESPE  is obtained from the diagonalization of the single-particle Hamiltonian,
\beqn 
h^p_q &=& \mathscr{F}^p_q + \sum_{rs}\mathscr{V}^{pr}_{qs}\rho^r_s\nonumber\\
&=&\bar{t}^p_q + \sum_{rs}\bar{v}^{pr}_{qs}\rho^r_s
+\dfrac{1}{4}\sum_{rstu} w^{prt}_{qsu} \gamma^{rt}_{su} 
+\sum_{rstu}w^{prt}_{qsu} \gamma^r_s(\rho^t_u-\gamma^t_u),\nonumber\\
\eeqn
where $\gamma^t_u$ is the one-body density of the correlated state, and $\rho^r_s$ is the  one-body density of mean-field state $\ket{\Phi^{\rm (OA)}_\kappa(\mathbf{q})}$ defined by 
\beq 
\rho^r_s \equiv\bra{\Phi^{\rm (OA)}_\kappa(\mathbf{q})}a^\dagger_r a_s\ket{\Phi^{\rm (OA)}_\kappa(\mathbf{q})}.
\eeq 
It is shown in Fig.~\ref{fig:SPE_interaction_A20} that the neutron partner states with the same value of $|m|$, related by the time-reversal operator, are not degenerate in the HFB states for \nuclide[21]{Ne} with odd-number parity. A comparison is made between the ESPEs obtained by the effective Hamiltonians {\tt magic-Ne20} and {\tt magic-Ne22}. The lifting of Kramers's degeneracy in the HFB states for \nuclide[21]{Ne} results in non-degeneracy among time-reversal states with identical values of $|m|$. For clarity, only the energy of one of the time-reversal states with $m>0$ is depicted in Fig.~\ref{fig:SPE_interaction_A20_A22}. It is observed that the ESPEs from the two effective Hamiltonians are difficult to distinguish.

\begin{figure}[]
\centering
\includegraphics[width=12.5 cm]{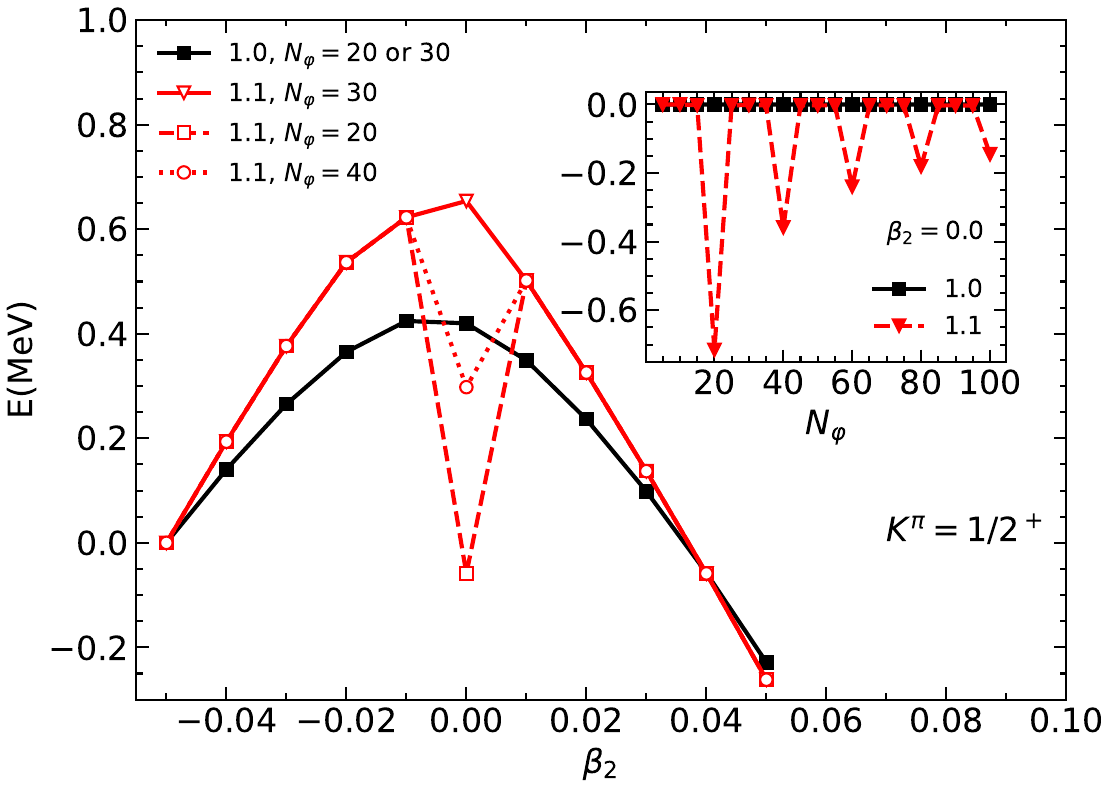}
\caption{The energies of PNP-HFB (VAPNP) states for \nuclide[21]{Ne}  with $K^\pi=1/2^+$ as a function of   quadrupole deformation $\beta_2$, where the number $N_\varphi$ of meshpoint in the gauge angle $\varphi$ is chosen as $20, 30$, and 40, respectively. The results from the calculations by multiplying a factor of $1.1$ artificially to the two-body interaction matrix elements for the mixed field $\tilde\Gamma$ are given for comparison. The inset shows the energy of  spherical state normalized to the converged value as a function of $N_\varphi$.
}
\label{fig:singularity}
\end{figure}

\begin{figure}[]
\centering
\includegraphics[width=\columnwidth]{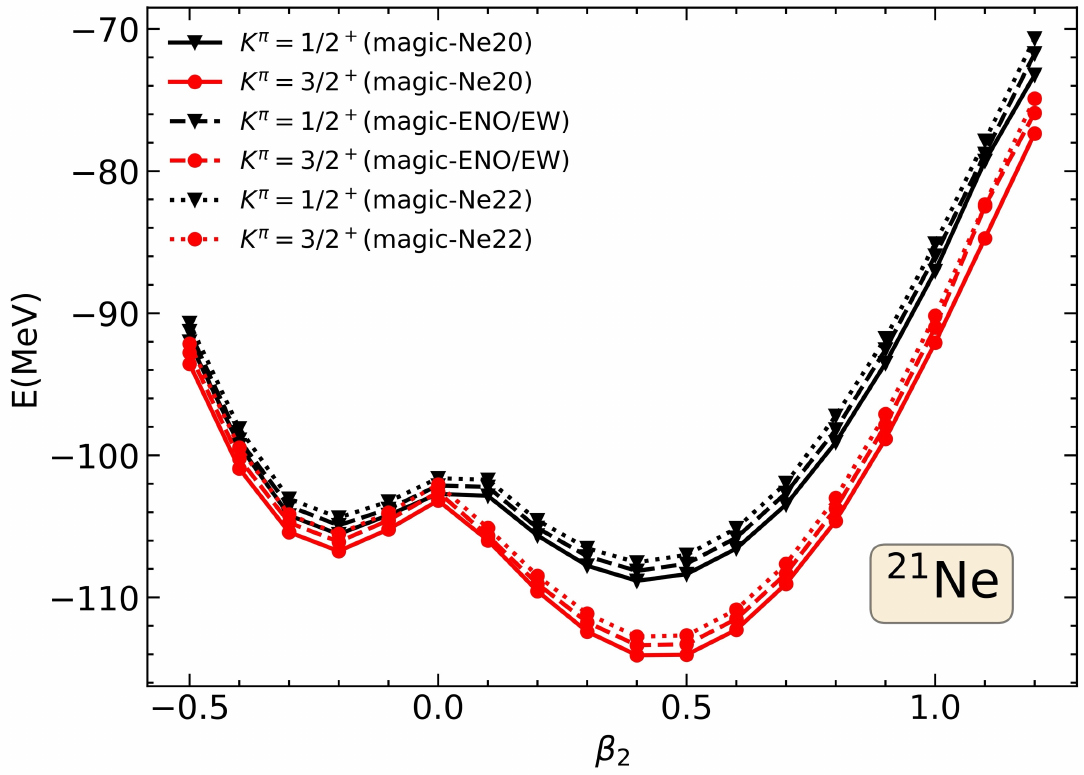}
\caption{ The energies of mean-field states $\ket{\Phi^{\rm (OA)}_\kappa(\mathbf{q})}$ for \nuclide[21]{Ne} with $K^\pi=1/2^+, 3/2^+$ as a function of intrinsic quadrupole deformation $\beta_2$ from the PNP-HFB (VAPNP) calculation using the three effective Hamiltonians. }
\label{fig:PES_interaction_A20_A22}
\end{figure}

Before presenting the projected energy curves with different angular momenta, we examine the issues of singularity and finite steps found in the MR-EDF~\cite{Bender:2009PRC,Duguet:2009PRC}. Figure~\ref{fig:singularity} displays the energies (normalized to the converged values) of PNP-HFB states for \nuclide[21]{Ne} with $K^\pi=1/2^+$ and quadrupole deformation $\beta_2=0.0$, as a function of the number $N_\varphi$ of meshpoints in the gauge angle $\varphi$. The Fomenko expansion method~\cite{Fomenko:1970JPA} is used for the particle-number projection, where the $k$-th gauge angle $\varphi_k$ is chosen as $2\pi(k/N_\varphi)$. It is observed that the energy remains constant for $N_\varphi \geq 5$, regardless of whether $N_\varphi$ is an even or odd number. For comparison, we also show the results from calculations by artificially multiplying a factor of $1.1$ to the two-body interaction matrix elements $\mathscr{V}$ for the mixed particle-hole field $\tilde \Gamma$. In this case, dips are indeed observed at $N_\varphi=20, 40, 60, \ldots$, corresponding to the situation where the gauge angle $\varphi_k=\pi/2$ is chosen at the meshpoints with $k=5, 10, 15, \ldots$, respectively. It demonstrates numerically that one should use the same interaction matrix elements for both the particle-hole and particle-particle channels in which case one is free of  the problem of singularity.

\begin{figure}[]
 \centering
 \includegraphics[width=0.45\columnwidth]{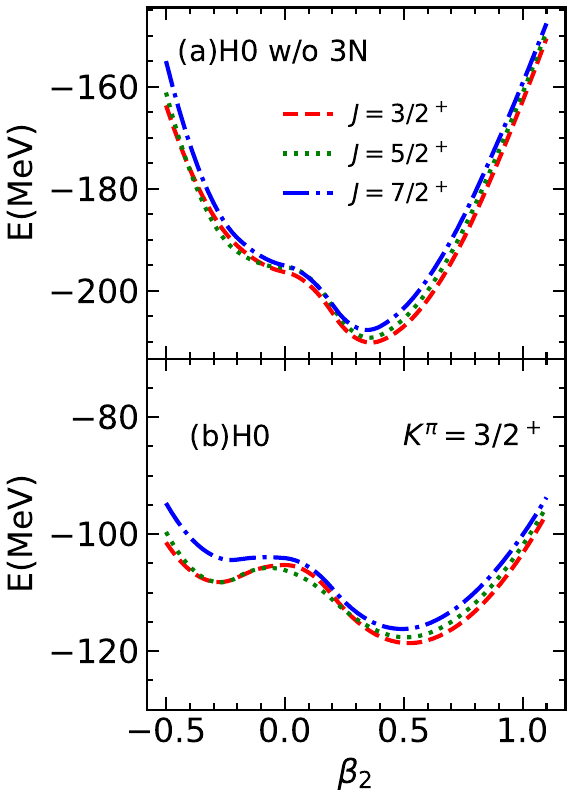} 
 \includegraphics[width=0.45\columnwidth]{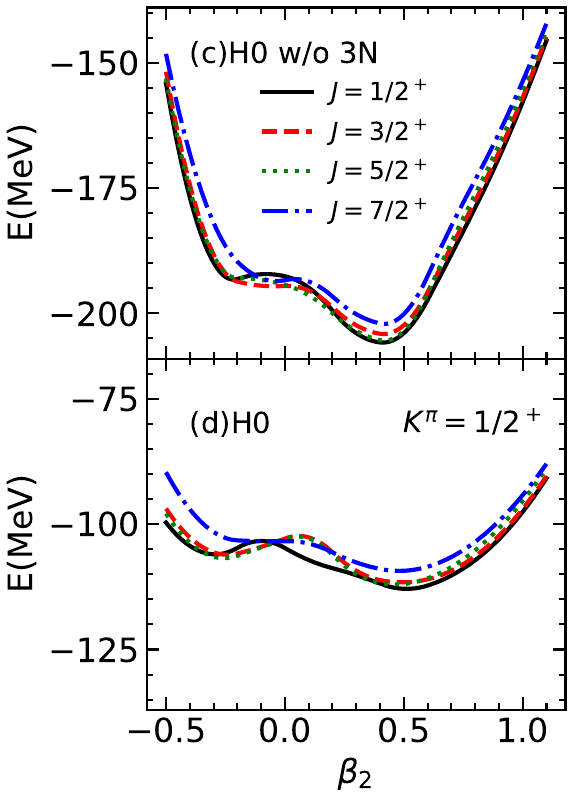} 
 \caption{The energies of states with projection onto particle numbers ($N=11, Z=10$) and spin-parity $J^\pi=3/2^+, 5/2^+$ and $7/2^+$ for $^{21}$Ne with quantum numbers $K^\pi=3/2^+$ (left panels) and $K^\pi=1/2^+$ (right panels) as a function of the quadrupole deformation parameter $\beta_2$ by the Hamiltonians {\tt magic-Ne20} (b, d) and {\tt magic-Ne20 (w/o 3N)} (a, c).  }
 \label{fig:Projcted_PES}
 \end{figure}

Figure~\ref{fig:PES_interaction_A20_A22} displays the energy curves of particle-number projected HFB states for \nuclide[21]{Ne} with $K^\pi=3/2^+$ and $1/2^+$, respectively. The HFB wave functions are obtained from the PNP-HFB (VAPNP) calculations using the Hamiltonian $\hat {\cal H}_0$, with the $3N$ interaction normal-ordered with respect to the references of \nuclide[20]{Ne}, \nuclide[22]{Ne}, and their ensemble with equal weights, respectively. It can be observed that the global energy minima of all three curves are located in prolate states with quadrupole deformation $\beta_2$ between 0.4 and 0.5. The configurations with $K^\pi=3/2^+$ are globally lower than those with $K^\pi=1/2^+$. Furthermore, the configurations based on different Hamiltonians are systematically shifted from each other in energy by less than one MeV.

Figure~\ref{fig:Projcted_PES} displays the energies of states with projection onto correct particle numbers and $J^\pi=3/2^+$, $5/2^+$, and $7/2^+$ for $^{21}$Ne with $K^\pi=3/2^+$ and $K^\pi=1/2^+$, respectively. The effective Hamiltonians used are {\tt H0} with and without the $3N$ interaction. It is shown that the quadrupole deformation parameter $\beta_2$ of the prolate energy-minimal state by the {\tt H0 (w/o 3N)} is smaller than the other two cases. Additionally, the energy curve with the increase of $\beta_2$ is also stiffer than that with the $3N$ interaction.  In other words, the 3N interaction helps the development of quadrupole collectivity in \nuclide[21]{Ne}.

\begin{figure}[]
\centering
\includegraphics[width=\columnwidth]{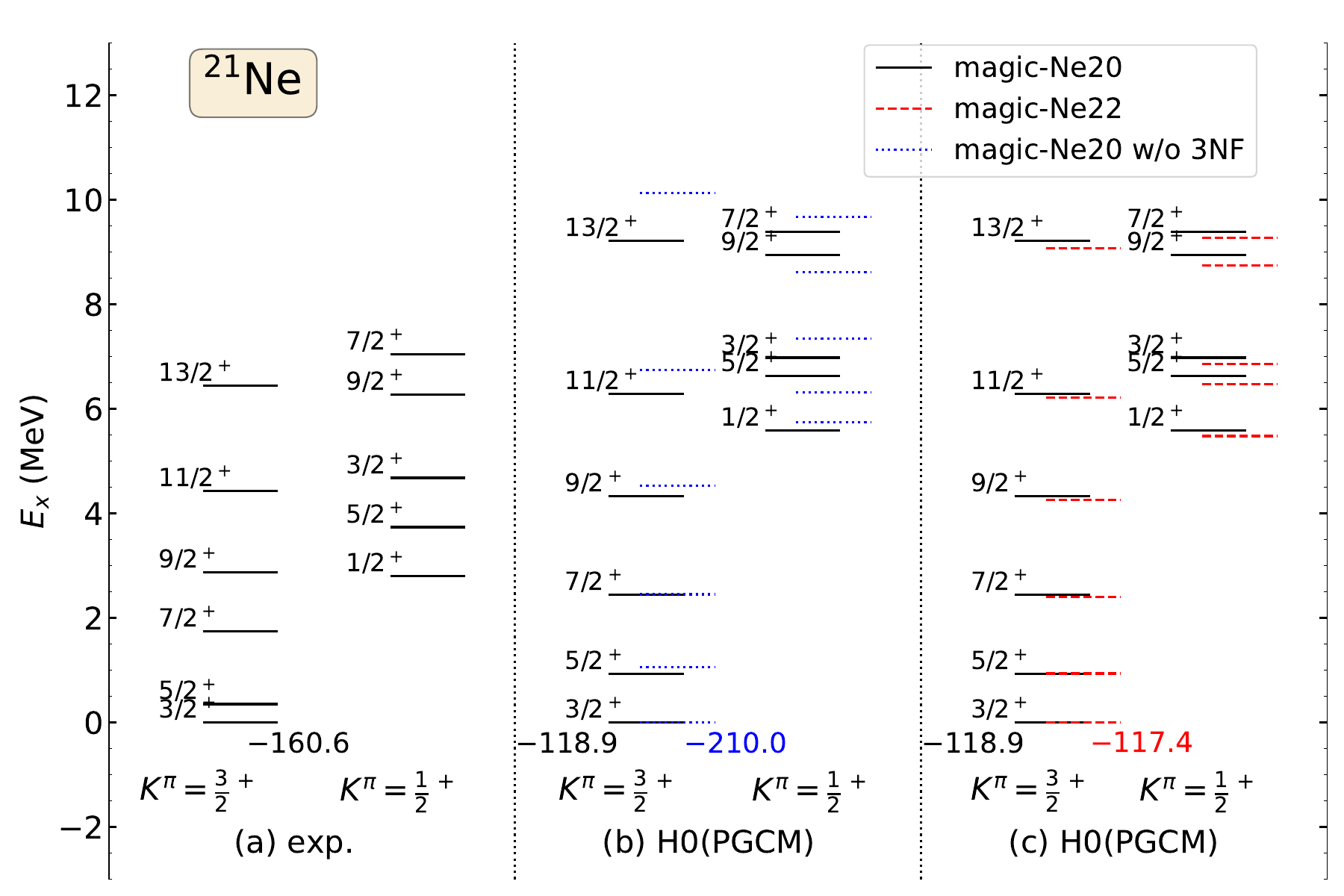}
\caption{The energy spectra of low-lying states in \nuclide[21]{Ne} with $K^\pi=3/2^+$ and $1/2^+$.  The corresponding data from Ref.~\cite{Data4Ne21_Firestone2015} are shown in (a). The results by the Hamiltonians {\tt magic-Ne20} and {\tt magic-Ne20 (w/o 3N)}  are displayed in (b). The results by the Hamiltonians {\tt magic-Ne20}  and {\tt magic-Ne22} are compared in (c).  }
\label{fig:spectra}
\end{figure}   

\begin{figure}[]
 \centering
 \includegraphics[width=0.45\columnwidth]{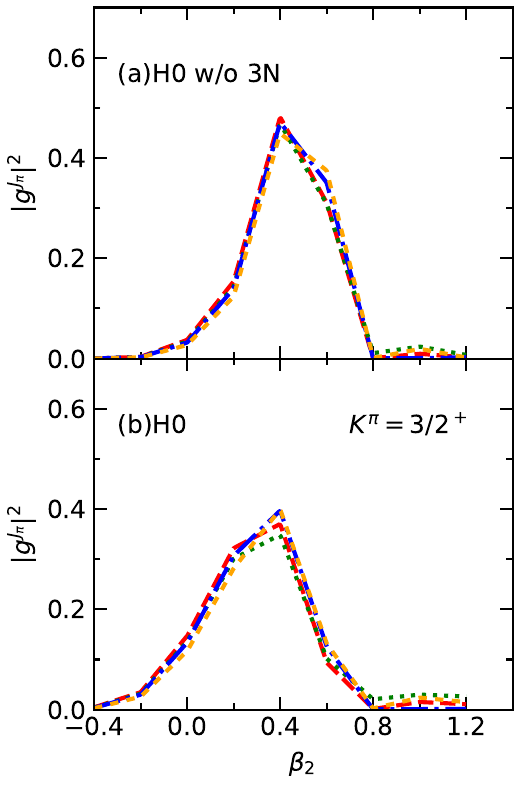} 
 \includegraphics[width=0.45\columnwidth]{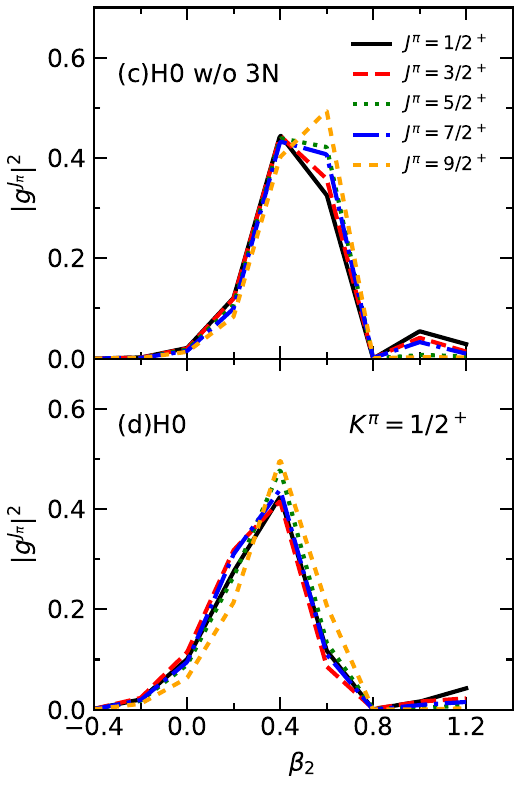} 
 \caption{The distribution of collective wave functions $|g^{J\pi}_\alpha|^2$, defined in (\ref{eq:coll_wf}), as a function of quadrupole deformation $\beta_2$ for the low-lying states of $^{21}$Ne with $K^\pi=3/2^+$ (left panels) and $K^\pi=1/2^+$ (right panels), respectively. The energy of the ground-state in each case is also provided.  }
 \label{fig:GCM-wfs}
 \end{figure}

Figure~\ref{fig:spectra} shows a comparison of the energy spectra for \nuclide[21]{Ne} from configuration-mixing calculations with different Hamiltonians. The states with the same $K^\pi$ are grouped into the same column. The main features of the two bands with $K^\pi=3/2^+$ and $1/2^+$ are reproduced, although the excitation energies of the states belonging to the $1/2^+$ band are systematically overestimated. The mixing of quasiparticle excitation configurations is expected to lower the entire $K^\pi=1/2^+$ band. In Fig.\ref{fig:spectra}(c), one can observe that the energy spectra from the  Hamiltonians {\tt magic-Ne20} and {\tt magic-Ne22} are very close to each other. The high-lying states from {\tt magic-Ne22} are slightly lower than those from {\tt magic-Ne20}. In Fig.\ref{fig:spectra}(b), the energy spectra become more stretched when the $3N$ interaction is turned off.  

The collective wave functions of the low-lying states with different $J^\pi$, and $K^\pi=3/2^+$ and $1/2^+$, by the {\tt magic-Ne20} effective Hamiltonian, are displayed in Fig.~\ref{fig:GCM-wfs}. It is shown that in all cases, the wave functions are peaked around $\beta_2=0.4$ and do not change significantly with the increase of angular momentum, implying the stability of the shapes in the low-lying states.

\section{Conclusions}
\label{sec:summary}

We have extended PGCM for the low-lying states of \nuclide[21]{Ne} starting from a chiral two-nucleon-plus-three-nucleon interaction, and compared the results obtained using effective Hamiltonians derived with the three-nucleon interaction normal-ordered with three different reference states: spherical particle-number projected HFB states for \nuclide[20]{Ne}, \nuclide[22]{Ne}, and an ensemble of them with equal weights. The topology of the potential energy surfaces (PES) shows no significant differences among the results by the three effective Hamiltonians, even though the PESs exhibit a systematic energy shift of less than one MeV. The excitation energies of the low-lying states of \nuclide[21]{Ne} by the effective Hamiltonian based on the reference state of \nuclide[20]{Ne} are slightly larger  than those by the effective Hamiltonian of \nuclide[22]{Ne}.  Furthermore, we demonstrate that the inclusion of the three-nucleon interaction significantly impacts the low-lying states. Without it, the energy spectrum becomes stretched, and the quadrupole collectivity is notably reduced.

This study provides a solid basis to extend the framework of IM-GCM~\cite{Yao:2018PRC,Yao:2020PRL}, namely, the combination of PGCM with ab initio method of multi-reference in-medium similarity renormalization group (MR-IMSRG)~\cite{Hergert:2016PR}, for the low-lying states of odd-mass nuclei based on consistently-evolved operators. The results of this study will be published elsewhere, separately.

\section*{Acknowledgments} 
This research was funded in part by the National Natural Science Foundation of China (Grant Nos.  12375119, 12141501, and 12005804), Guangdong Basic and Applied Basic Research Foundation (2023A1515010936) and the Fundamental Research Funds for the Central Universities, Sun Yat-sen University. H.H. was funded by  the U.S. Department of Energy, Office of Science, Office of Nuclear Physics DE-SC0017887, DE-SC0023516, as
well as DE-SC0018083, DE-SC0023175 (SciDAC NUCLEI Collaboration).

 \bibliographystyle{apsrev4-1} 

\end{document}